\documentclass[journal,twocolumn]{IEEEtran}
\usepackage{stmaryrd}
\usepackage{amsfonts}
\usepackage{amsmath}
\usepackage{amssymb}
\usepackage{cite}
\usepackage{graphicx}
\usepackage{subfigure}
\usepackage{amsthm}
\usepackage{algorithmic}
\usepackage{algorithm}
\usepackage{stfloats}

\newtheorem{Theorem}{Theorem}
\newtheorem{Definition}{Definition}
\newtheorem{Lemma}[Theorem]{Lemma}

\newtheorem{Example}{Example}

\begin{document}
%

\title{Streaming Algorithms for Optimal Generation of Random Bits}

\author{Hongchao~Zhou,
        and~Jehoshua~Bruck,~\IEEEmembership{Fellow,~IEEE}
\thanks{Hongchao~Zhou and Jehoshua~Bruck are with the Department
of Electrical Engineering, California Institute of Technology, Pasadena,
CA 91125, USA, e-mail: hzhou@caltech.edu; bruck@caltech.edu.}
\thanks{This work was supported in part by the NSF Expeditions in Computing
Program under grant CCF-0832824.}
}


\maketitle%
\begin{abstract}
Generating random bits from a source of biased coins (the biased is unknown) is a classical question that was originally studied by von Neumann.
There are a number of known algorithms that have asymptotically optimal information efficiency, namely, the
expected number of generated random bits per input bit is asymptotically close to the entropy of the source.
However, only the original von Neumann algorithm has a `streaming property' - it operates on a single input
bit at a time and it generates random bits when possible, alas, it does not have an optimal information efficiency.

The main contribution of this paper is an algorithm that generates random bit streams from biased coins, uses bounded space and runs in expected linear time. As the size of the allotted space increases, the algorithm approaches the information-theoretic upper bound on efficiency.
In addition, we discuss how to extend this algorithm to generate random bit streams from $m$-sided dice or correlated sources such as Markov chains.
\end{abstract}

\begin{IEEEkeywords}
Random Number Generation, Biased Coins, Markov Chains, Streams.
\end{IEEEkeywords}

\IEEEpeerreviewmaketitle

\section{Introduction}

\IEEEPARstart{T}{he} question of generating random bits from a source of biased coins dates back to von Neumann \cite{Neumann1951} who observed that
when one focuses on a pair of coin tosses, the events  HT and TH have the same probability (H is for `head' and T is for `tail') of being generated; hence, HT produces the output symbol $1$ and TH produces the output symbol $0$. The other two possible events, namely, HH and TT, are ignored, namely, they do not produce any output symbols. However, von Neumann's algorithm is not optimal in terms of the number of random bits that
are generated. This problem was solved, specifically, given a fixed number of biased coin tosses with unknown probability, it is well known how to generate random bits with asymptotically optimal efficiency, namely, the expected number of unbiased random bits generated per coin toss is asymptotically equal to the entropy of the biased coin \cite{Pae2005, Elias1972,Peres1992,Ryabko2000}. However, these solutions, including Elias's algorithm and Peres's algorithm, can generate random bits only after receiving the complete input sequence (or a fixed number of input bits),
and the number of random bits generated is a random variable.

We consider the setup of generating a ``stream" of random bits; that is, whenever random bits are required, the algorithm reads
new coin tosses and generates random bits dynamically. Our new streaming algorithm is more efficient (in the number of input bits, memory and time) for producing the required number of random bits and is a better choice for implementation in practical systems. We notice that von Neumann scheme is the one which is able to generate a stream of random bits, but its efficiency is far from optimal. Our goal is to modify
this scheme such that it can achieve the information-theoretic upper bound on efficiency. Specifically, we would like to construct a function $f: \{\textrm{H},\textrm{T}\}^*\rightarrow\{0,1\}^*$ which satisfies the following conditions:

\begin{itemize}
 \item $f$ generates a stream. For any two sequences of coin tosses $x,y\in \{\textrm{H},\textrm{T}\}^*$, $f(x)$ is a prefix of $f(xy)$.
   \item $f$ generates random bits. Let $X_k\in \{0,1\}^*$ be the sequence of coin tosses inducing $k$ bits; that is, $|f(X_k)|\geq k$ but
   for any strict prefix $X$ of $X_k$, $|f(X)|\leq k$.  Then the first $k$ bits of $f(X_k)$ are independent and unbiased.
   \item $f$ has asymptotically optimal efficiency. That is, for any $k>0$,
   $$\frac{E[|X_k|]}{k}\rightarrow\frac{1}{H(p)} $$
   as $k\rightarrow \infty$,
   where $H(p)$ is the entropy of the biased coin \cite{Cover2006}.
\end{itemize}

We note that the von Neumann scheme uses only $3$ states, i.e., a symbol in $\{\phi,\textrm{H}, \textrm{T}\}$, for storing state information.
For example, the output bit is $1$ if and only if the current state is H and the input symbol is T. In this case, the new state is $\phi$.
Similarly, the output bit is $0$ if and only if the current state is T and the input symbol is H. In this case, the new state is $\phi$.
Our approach for generalizing von Neumann's scheme is by increasing the memory (or state) of our algorithm such that we do not
lose information that might be useful for generating future random bits. We represent the state information as a binary tree, called status tree,
in which each node is labeled by a symbol in $\{\phi, \textrm{H}, \textrm{T}, 0, 1\}$.  When a source symbol (a coin toss) is received,
we modify the status tree based on certain simple rules and generate random bits in a dynamic way. This is the key idea in our
algorithm; we call this approach the random-stream algorithm. In some sense, the random-stream algorithm is the streaming version of Peres's algorithm. We show that this algorithm satisfies all three conditions above, namely, it can generate a stream of random bits with
asymptotically optimal efficiency. In practice, we can reduce the space size by limiting the depth of the status tree. We will demonstrate that as the depth of the status tree increases, the efficiency of the algorithm quickly converges to the information-theoretic upper bound.

An extension of the question is to generate random bits or random-bit streams from an arbitrary Markov chain with unknown transition probabilities. This problem was first studied by Samuelson \cite{Samuelson1968}, and his algorithm was later improved by Blum \cite{Blum1986}. Recently, we proposed the first known algorithm that runs in expected linear time and achieves the information-theoretic upper bound on efficiency \cite{Zhou2012Markov}. In this paper, we briefly introduce the techniques of generating random-bit streams from Markov chains.

The rest of the paper is organized as follows.
Section \ref{Stream_section_stream} presents our key result, the random-stream algorithm that generates random bit streams from arbitrary biased coins and achieves the information-theoretic upper bound on efficiency.
In Section \ref{Stream_section_generalization}, we generalize the random-stream algorithm to generate random bit streams from a source of a larger alphabet.
An extension for Markov chains is provided in Section \ref{Stream_section_Markov}, followed by the concluding remarks.

\section{The Random-Stream Algorithm}
\label{Stream_section_stream}

\subsection{Description}

Many algorithms have been proposed for efficiently generating random bits from a fixed number of coins tosses, including Elias's algorithm and Peres's algorithm. However, in these algorithms, the input bits can be processed only after all of them have been received, and the number of random bits generated cannot be controlled. In this section, we focus on deriving
a new algorithm, the \emph{random-stream algorithm}, that generates a stream of random bits from an arbitrary biased-coin source and achieves the information-theoretic upper bound on efficiency. Given an application that requires random bits, the random-stream algorithm can generate random bits dynamically based on requests from the application.

While von Neumann's scheme can generate a stream of random bits from an arbitrary biased coin, its efficiency is far from being optimal. The main reason is that it uses minimal state information, recorded by a symbol of alphabet size three in $\{\phi, \textrm{H},\textrm{T}\}$. The key idea in our algorithm is to create a binary tree for storing the state information, called a status tree. A node in the status tree stores a symbol in $\{\phi,\textrm{H}, \textrm{T}, 0, 1\}$.
The following procedure shows how the status tree is created and is dynamically updated in response to arriving input bits. At the beginning, the tree has only a single root node labeled as $\phi$. When reading a coin toss from the source,
we modify the status tree based on certain rules. For each node in the status tree, if it receives a message (H or T), we do operations on the node.  Meanwhile, this node may pass some new messages to its children.  Iteratively, we can process the status tree until no more messages are generated. Specifically, let $u$ be a node in the tree. Assume the label
of $u$ is $x \in \{\phi,\textrm{H}, \textrm{T}, 1, 0\}$ and it receives a symbol $y\in\{\textrm{H}, \textrm{T}\}$ from its parent node (or from the source if $u$ is
the root node). Depending on the values of x and y, we do the following operations on node $u$.
\begin{enumerate}
  \item When $x=\phi$, set $x=y$.
  \item When $x=1$ or $0$, output $x$ and set $x=y$.
  \item When $x=\textrm{H}$ or $\textrm{T}$, we first check whether $u$ has children. If it does not have, we create two children with label $\phi$ for it.
Let $u_l$ and $u_r$ denote the two children of $u$.
    \begin{itemize}
  \item If $xy=\textrm{HH}$, we set $x=\phi$, then pass a symbol $\textrm{T}$ to $u_l$ and a symbol $\textrm{H}$ to $u_r$.
  \item If $xy=\textrm{TT}$, we set $x=\phi$, then pass a symbol $\textrm{T}$ to $u_l$ and a symbol $\textrm{T}$ to $u_r$.
  \item If $xy=\textrm{HT}$, we set $x=1$, then pass a symbol $\textrm{H}$ to $u_l$.
  \item If $xy=\textrm{TH}$, we set $x=0$, then pass a symbol $\textrm{H}$ to $u_l$.
\end{itemize}
We see that the node $u$ passes a symbol $x+y\mod2$ to its left child and if $x=y$ it passes a symbol $x$ to its right child.
\end{enumerate}

Note that the timing is crucial that we output a node's label (when it is $1$ or $0$) only after it receives the next symbol from
its parent or from the source. This is different from von Neumann's scheme where a $1$ or a $0$ is generated immediately without waiting for the next symbol. If we only consider the output of the root node in the status tree, then it is similar to von Neumann's scheme. And the other nodes correspond to the information discarded by von Neumann's scheme. In some sense, the random-stream algorithm can be treated as a ``stream" version of Peres's algorithm.  The following example is constructed for the purpose of demonstration.

\begin{figure*}[!t]
\centering
\includegraphics[width=4.5in]{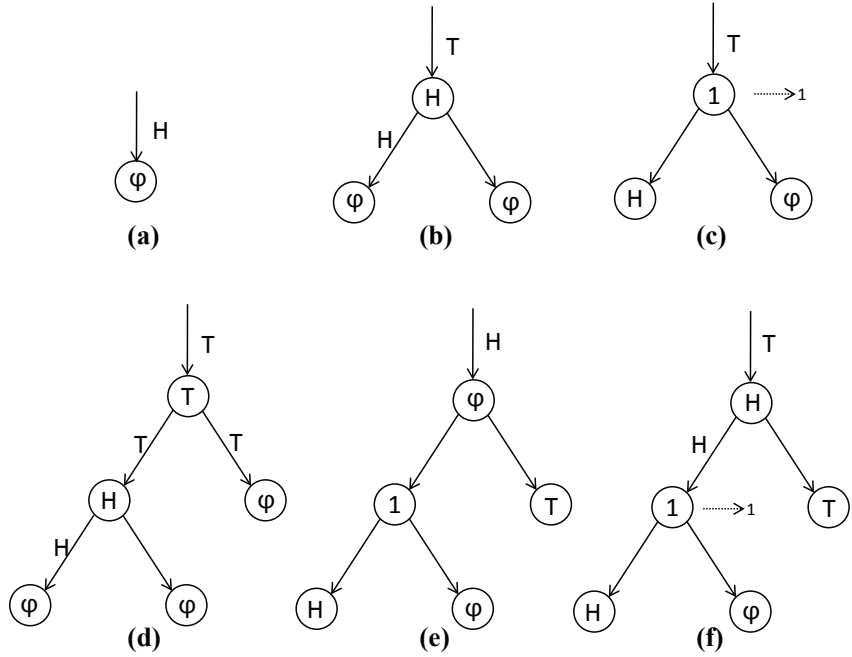}
\caption{An instance for generating $2$ random bits using the random-stream algorithm.}
\label{fig_peresscheme}
\end{figure*}

\begin{Example} \label{example1_1} Assume we have a biased coin and our randomized application requires $2$ random bits. Fig.~\ref{fig_peresscheme} illustrates how the random-stream algorithm works when the incoming stream is HTTTHT... In this figure, we can see the changes of the status tree and the messages (symbols) passed throughout the tree for each step. We see that the output stream is
$11...$
\end{Example}

\begin{Lemma}\label{Stream_lemma1_1}
Let $X$ be the current input sequence and let $\mathcal{T}$ be the current status tree. Given $\mathcal{T}$ and the bits generated by each node
in $\mathcal{T}$, we can reconstruct $X$ uniquely.
\end{Lemma}

\proof Let us prove this lemma by induction. If the maximum depth of the status tree is $0$, it has only a single node. In this case,
$X$ is exactly the label on the single node. Hence the conclusion is trivial. Now we show that if the conclusion holds for all status trees with maximum depth at most $k$, then it also holds for all status trees with maximum depth $k+1$.

Given a status tree $\mathcal{T}$ with maximum depth $k+1$, we let $Y\in \{0,1\}^*$ denote the binary sequence generated by the root node, and $L, R\in \{\textrm{H},\textrm{T}\}^*$ are the sequences of symbols received by its left child and right child. If the label of the root node is in $\{0,1\}$, we add it to $Y$.
According
to the random-stream algorithm, it is easy to get that
$$|L|=|Y|+|R|.$$

Based on our assumption, $L, R$ can be constructed
from the left and right subtrees and the bits generated by each node in the subtree since their depths are at most $k$. We show that once $L, R, Y$ satisfy the equality above,
the input sequence $X$ can be uniquely constructed from $L, R, Y$ and $\alpha$, where $\alpha$ is the label of the root node. The procedure is as follows: Let us start from an empty string for $X$ and read symbols from $L$ sequentially. If a symbol read from $L$ is $\textrm{H}$, we read a bit from $Y$. If this bit is $1$ we add $\textrm{HT}$ to $X$, otherwise we add $\textrm{TH}$ to $X$. If a symbol read from $L$ is $\textrm{T}$, we read a symbol ($\textrm{H}$ or $\textrm{T}$) from $R$. If this symbol is $\textrm{H}$ we add $\textrm{HH}$ to $X$, otherwise we add $\textrm{TT}$ to $X$.

After reading all the elements in $L, R$ and $Y$, the length of the resulting input sequence is $2|L|$. Now, we add $\alpha$ to the resulting input sequence if $\alpha\in \{\textrm{H},\textrm{T}\}$. This leads to the final sequence $X$, which is unique.
\hfill\QED

\begin{Example} Let us consider the status tree in Fig.~\ref{fig_peresscheme}(f). And we know that the root node generates $1$ and
the first node in the second level generates $1$. We can have the following conclusions iteratively.
\begin{itemize}
  \item In the third level, the symbols received by the node with label H are H, and the node with label $\phi$ does not receive any symbols.
  \item In the second level, the symbols received by the node with label $1$ are HTH, and the symbols received by the node with label T are T.
  \item For the root node, the symbols received are HTTTHT, which accords with Example \ref{example1_1}.
\end{itemize}
\end{Example}

Let $f:\{\textrm{H},\textrm{T}\}^*\rightarrow\{0,1\}^*$ be the function of the random-stream algorithm. We show that this function satisfies all
the three conditions described in the introduction. It is easy to see that the first condition holds, i.e., for any two sequences $x,y\in \{\textrm{H},\textrm{T}\}^*$,
$f(x)$ is a prefix of $f(xy)$, hence it generates streams. The following two theorems indicate that $f$ also satisfies the other two conditions.

\begin{Theorem} \label{Stream_theorem1}Given a source of biased coin with unknown probability, the random-stream algorithm generates a stream of random bits, i.e.,
for any $k>0$, if we stop running the algorithm after generating $k$ bits then these $k$ bits are independent and unbiased.
\end{Theorem}

Let $S_Y$ with $Y\in \{0,1\}^k$ denote the set consisting of all the binary sequences yielding $Y$. Here,
we say that a binary sequence $X$ \emph{yields} $Y$ if and only if $X[1:|X|-1]$ (the prefix of $X$ with length $|X|-1$) generates a sequence shorter than $Y$ and
$X$ generates a sequence with $Y$ as a prefix (including $Y$ itself).  To prove that the algorithm can generate
random-bit streams, we show that for any distinct binary sequences $Y_1, Y_2\in \{0,1\}^k$, the elements in $S_{Y_1}$ and those in $S_{Y_2}$ are
one-to-one mapping. The detailed proof is given in Subsection \ref{section_proof1}.

\begin{Theorem} \label{Stream_theorem2}
Given a biased coin with probability $p$ being H, let $n$ be the number of coin tosses required for generating $k$ random bits in the random-stream algorithm, then
$$\lim_{k\rightarrow\infty}\frac{E[n]}{k}= \frac{1}{H(p)}.$$
\end{Theorem}

The proof of Theorem \ref{Stream_theorem2} is based on the fact that the random-stream algorithm is as efficient as Peres's algorithm. The difference is
that in Peres's algorithm the input length is fixed and the output length is variable. But in
the random-stream algorithm the output length is fixed and the input length is variable. So the key of the proof is to connect these two cases. The detailed proof is given in Subsection \ref{section_proof2}.

So far, we can conclude that the random-stream algorithm can generate a stream of random bits from an arbitrary biased coin with asymptotically optimal efficiency. However, the size of the binary tree increases as the number of input coin tosses increases. The longest path of the tree
is the left-most path, in which each node passes one message to the next node when it receives two messages from its previous node.
Hence, the maximum  depth of the tree is $\log_2 n$ for $n$ input bits.
 This linear increase in space is a practical challenge. Our observation is that we can control the size of the space by limiting the maximum depth of the tree -- if
a node's depth reaches a certain threshold, it will stop creating new leaves. We can prove that this method correctly generates
a stream of random bits from an arbitrary biased coin. We call this method the random-stream algorithm with maximum depth $d$.

\begin{Theorem} \label{Stream_theorem7} Given a source of a biased coin with unknown probability, the random-stream algorithm with maximum depth $d$ generates a stream of random bits, i.e.,
for any $k>0$, if we stop running the algorithm after generating $k$ bits then these $k$ bits are independent and unbiased.
\end{Theorem}

The proof of Theorem \ref{Stream_theorem7} is a simple modification of the proof of Theorem \ref{Stream_theorem1}, given in Subsection \ref{section_proof3}. In order to save memory space,
we need to reduce the efficiency. Fortunately, as the maximum depth increases, the efficiency of this method can quickly converge to the theoretical limit.

\begin{Example} When the maximum depth of the tree is $0$ (it has only the root node), then the algorithm is approximately von Neumann's scheme.
The expected number of coin tosses required per random bit is
$$\frac{1}{pq}$$
 asymptotically, where $q=1-p$ and $p$ is the probability for the  biased coin being H.
\end{Example}

\begin{Example} \label{Stream_theorem3} When the maximum depth of the tree is $1$, the expected number of coin tosses required per random bit is
$$\frac{1}{pq+\frac{1}{2}(p^2+q^2)(2pq)+\frac{1}{2}(p^2+q^2)\frac{p^2q^2}{(p^2+q^2)^2}}$$
asymptotically, where $q=1-p$ and $p$ is the probability for the biased coin being H.
\end{Example}

Generally, if the maximum depth of the tree is $d$, then we can calculate the efficiency of the random-stream algorithm by iteration in the following way:

\begin{Theorem}\label{stream_theorem4} When the maximum depth of the tree is $d$ and the probability of the biased coin is $p$ of being H, the expected number of coin tosses
required per random bit is
$$\frac{1}{\rho_d(p)}$$ asymptotically, where $\rho_d(p)$ can be obtained by iterating
\begin{equation}\rho_d(p)=pq+\frac{1}{2}\rho_{d-1}(p^2+q^2)+\frac{1}{2}(p^2+q^2)\rho_{d-1}(\frac{p^2}{p^2+q^2})\label{stream_equ_efficiencydepthd}\end{equation}
with $q=1-p$ and $\rho_0(p)=pq$.
\end{Theorem}

Theorem \ref{stream_theorem4} shows that the efficiency of a random-stream algorithm with maximum depth $d$ can be easily calculated by iteration.  One thing that we can claim is,
$$\lim_{d\rightarrow\infty}\rho_d(p)= H(p).$$
However, it is difficult to get an explicit expression for $\rho_d(p)$ when $d$ is finite.
As $d$ increases, the convergence rate of $\rho_d(p)$ depends on the value of $p$. The following extreme case implies that $\rho_d(p)$ can converge to $H(p)$ very quickly.

\begin{table}[!t]
  \centering
  \caption{The expected number of coin tosses required per random bit for different probability $p$ and different maximum depths}
   \renewcommand{\arraystretch}{1.5}
\begin{tabular}{|c|c|c|c|c|c|}
  \hline
  maximum depth & p=0.1 & p=0.2 & p=0.3 & p=0.4  & p=0.5\\
  \hline
   0 & 11.1111  &  6.2500  &  4.7619  &  4.1667  &  4.0000\\
   1 &  5.9263 &   3.4768  &  2.7040  &  2.3799  &  2.2857\\
   2 &  4.2857 &   2.5816 &   2.0299 &   1.7990  &  1.7297\\
   3 & 3.5102  &  2.1484 &   1.7061  &  1.5190  &  1.4629\\
   4 & 3.0655 &   1.9023  &  1.5207  &  1.3596  &  1.3111\\
   5 & 2.7876  &  1.7480  &  1.4047  &  1.2598  &  1.2165\\
   7 & 2.4764  &  1.5745  &  1.2748  &  1.1485  &  1.1113\\
   10 & 2.2732  &  1.4619  &  1.1910  &  1.0772  &  1.0441\\
   15 & 2.1662  &  1.4033  &  1.1478  &  1.0408 &   1.0101\\
  $\infty$ & 2.1322   &   1.3852      &  1.1347  &  1.0299  &  1.0000  \\
  \hline
\end{tabular}
\label{table1}
\end{table}

\begin{Example} Let us consider the case that $p=\frac{1}{2}$. According to Equ. (\ref{stream_equ_efficiencydepthd}), we have
$$\rho_d(\frac{1}{2})=\frac{1}{4}+\frac{1}{2}\rho_{d-1}(\frac{1}{2})+\frac{1}{4}\rho_{d-1}(\frac{1}{2}),$$
where $\rho_0(\frac{1}{2})= \frac{1}{4}$. Based on this iterative relation, it can be obtained that
$$\rho_d(\frac{1}{2})=1-(\frac{3}{4})^{d+1}.$$
So when $p=\frac{1}{2}$, $\rho_d(p)$ can converge to  $H(p)=1$ very quickly as $d$ increases.
\end{Example}

\begin{figure}[!t]
\centering
\includegraphics[width=3.6in]{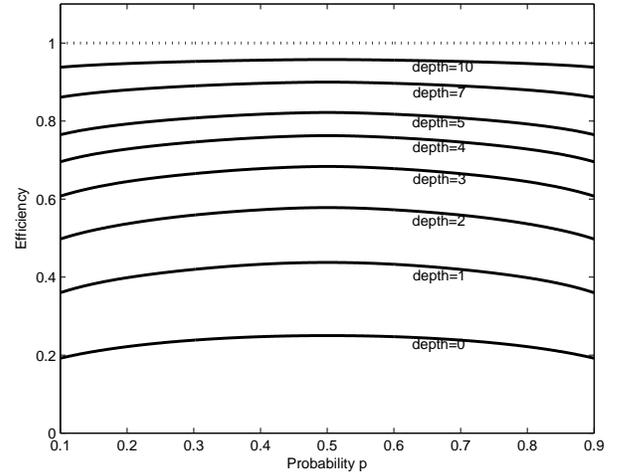}
\caption{The efficiency for different probability $p$ and different maximum depths.}
\label{fig_streamingefficiency}
\end{figure}

In Table \ref{table1}, we tabulate the expected number of coin tosses required per random bit in the random-stream algorithm with different maximum depths.
We see that as the maximum depth increases, the efficiency of the random-stream algorithm approaches the theoretical limitation quickly.
Let us consider the case of $p=0.3$ as an example. If the maximum depth is $0$, the random-stream algorithm is as efficient as von Neumann's scheme, which
requires expected $4.76$ coin tosses to generate one random bit. If the maximum depth is $7$, it requires only expected $1.27$ coin
tosses to generate one random bit. That is very close to the theoretical limitation $1.13$.
However, the space cost of the algorithm has an exponential dependence on the maximum depth. That requires us to balance
the efficiency and the space cost in real applications. Specifically, if we define efficiency as the ratio between
the theoretical lower bound and the real value of the expected number of coin tosses, then Fig. \ref{fig_streamingefficiency}
shows the relation between the efficiency and the maximum depth for different probability $p$.

\begin{table}[!t]
  \centering
   \caption{The expected time for processing a single input coin toss for different probability $p$ and different maximum depths}
   \renewcommand{\arraystretch}{1.5}
\begin{tabular}{|c|c|c|c|c|c|}
  \hline
  maximum depth & p=0.1 & p=0.2 & p=0.3 & p=0.4  & p=0.5\\
  \hline
   0 & 1.0000   & 1.0000  &  1.0000 &   1.0000  &  1.0000 \\
   1 & 1.9100   & 1.8400  &  1.7900  &  1.7600  &  1.7500 \\
   2 & 2.7413  &  2.5524  &  2.4202  &  2.3398  &  2.3125 \\
   3 & 3.5079  &  3.1650  &  2.9275  &  2.7840  &  2.7344 \\
   4 & 4.2230  &  3.6996 &   3.3414  &  3.1256  &  3.0508 \\
   5 & 4.8968  &  4.1739 &   3.6838  &  3.3901  &  3.2881 \\
   7 & 6.1540  &  4.9940  &  4.2188  &  3.7587  &  3.5995 \\
   10 & 7.9002  &  6.0309  &  4.8001  &  4.0783 &   3.8311 \\
   15 & 10.6458  &  7.5383 &   5.5215  &  4.3539 &   3.9599 \\
  \hline
\end{tabular}
 \label{table2}
\end{table}

Another property that we consider is the expected time for processing a single coin toss. Assume that it takes a single unit of time to process a message received at a node, then the expected time is exactly the expected number of messages that have been generated in the status tree (including the input coin toss itself). Table \ref{table2} shows the expected time for processing a single input bit when the input is infinitely long, implying the computational efficiency of the random-stream algorithm with limited depth. It can be proved that for an input generated by an arbitrary biased coin the expected time for processing a single coin toss is upper bounded by the maximum depth plus one (it is not a tight bound).

\subsection{Proof of Theorem \ref{Stream_theorem1}}
\label{section_proof1}

In this subsection, we prove Theorem \ref{Stream_theorem1}.

\begin{Lemma}\label{Stream_lemma1_2}
Let $\mathcal{T}$ be the status tree induced by $X_A\in \{H,T\}^*$, and let $k_1, k_2, ..., k_{|\mathcal{T}|}$ be the number of bits generated by the nodes in $\mathcal{T}$, where $|\mathcal{T}|$ is the number of nodes in $\mathcal{T}$.
Then for any $y_i\in \{0,1\}^{k_i}$ with $1\leq i\leq |\mathcal{T}|$, there exists an unique sequence $X_B\in \{\textrm{H},\textrm{T}\}^*$ such that it induces the same status tree $\mathcal{T}$, and the bits generated by the $i$th node in $\mathcal{T}$ is $y_i$. For such a sequence $X_B$, it is a permutation of $X_A$ with the same last element.
\end{Lemma}

\proof To prove this conclusion, we can apply the idea of Lemma \ref{Stream_lemma1_1}. It is obviously that if the maximum depth of $\mathcal{T}$ is zero, then
the conclusion is trivial. Assume that the conclusion holds for any status tree with maximum depth at most $k$, then we show that it also holds for any status tree with maximum depth $k+1$.

Given a status tree $\mathcal{T}$ with maximum depth $k+1$, we use $Y_A\in \{0,1\}^*$ to denote the binary sequence generated by the root node based on $X_A$, and $L_A, R_A$ to denote the sequences of symbols received by its left child and right child. According to our assumption, by flipping
the bits generated by the left subtree, we can construct an unique sequence $L_B\in \{\textrm{H},\textrm{T}\}^*$ uniquely such that $L_B$ is
a permutation of $L_A$ with the same last element. Similarly, for the right subtree, we have $R_B\in\{\textrm{H},\textrm{T}\}^*$ uniquely such that
$R_B$ is a permutation of $R_A$ with the same last element.

Assume that by flipping the bits generated by the root node, we get a binary sequence $Y_B$ such that $|Y_B|=|Y_A|$ (If the label $\alpha\in\{0,1\}$, we add it to $Y_A$ and $Y_B$), then
$$|L_B|=|Y_B|+|R_B|,$$
which implies that we can construct $X_B$ from $L_B, R_B, Y_B$ and the label $\alpha$ on the root node uniquely (according to the proof of the above lemma). Since the length of $X_B$ is uniquely determined by $|L_B|$ and $\alpha$, we can also conclude that $X_A$ and $X_B$
have the same length.

To see that $X_B$ is a permutation of $X_A$, we show that $X_B$ has the same number of $\textrm{H}$'s as $X_A$. Given a sequence $X\in \{\textrm{H},\textrm{T}\}^*$, let $w_\textrm{H}(X)$ denote the number of $\textrm{H}$'s in $X$. It is not hard to see that
$$w_\textrm{H}(X_A)=w_\textrm{H}(L_A)+w_\textrm{H}(R_A)+w_\textrm{H}(\alpha),$$
$$w_\textrm{H}(X_B)=w_\textrm{H}(L_B)+w_\textrm{H}(R_B)+w_\textrm{H}(\alpha),$$
where $w_\textrm{H}(L_A)=w_\textrm{H}(L_B)$ and $w_\textrm{H}(R_A)=w_\textrm{H}(R_B)$ due to our assumption. Hence $w_\textrm{H}(X_A)=w_\textrm{H}(X_B)$ and $X_B$ is a permutation of $X_A$.

Finally, we would like to see that $X_A$ and $X_B$ have the same last element. That is because if $\alpha\in \{\textrm{H},\textrm{T}\}$, then both $X_A$ and $X_B$ end with $\alpha$. If $\alpha\in \{\phi, 0, 1\}$,
the last element of $X_B$ depends on the last element of $L_B$, the last element of $R_B$, and $\alpha$. Our assumption gives that
$L_B$ has the same element as $L_A$ and $R_B$ has the same element as $R_A$. So we can conclude that $X_A$ and $X_B$ have the same last element.

This completes the proof.\hfill\QED

\begin{figure}[!t]
\centering
\includegraphics[width=3.4in]{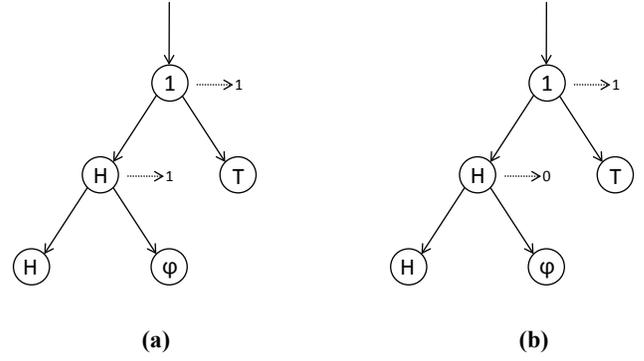}
\caption{An example for demonstrating Lemma \ref{Stream_lemma1_2}, where the input sequence for (a) is HTTTHT, and the
input sequence for (b) is TTHTHT.}
\label{fig_perestree}
\end{figure}

\begin{Example} The status tree of a sequence HTTTHT is given by Fig. \ref{fig_perestree}(a). If we flip the second bit $1$ into $0$, see
 Fig. \ref{fig_perestree}(b), we can construct a sequence of coin tosses , which is TTHTHT.
\end{Example}

Now, we define an equivalence relation on $\{\textrm{H},\textrm{T}\}^*$.

\begin{Definition}\label{definition_1}
Let $\mathcal{T}_A$ be the status tree of $X_A$ and $\mathcal{T}_B$ be the status tree of $X_B$. Two sequences $X_A, X_B\in \{H,T\}^*$ are equivalent
 denoted by $X_A\equiv X_B$ if and only if $\mathcal{T}_A=\mathcal{T}_B$, and for each pair of nodes $(u,v)$ with $u\in T_A$ and $v\in T_B$
 at the same position they generate the same number of bits.
\end{Definition}

Let $S_Y$ with $Y\in \{0,1\}^k$ denote the set consisting of all the binary sequences yielding $Y$. Here,
we say that a binary sequence $X$ \emph{yields} $Y$ if and only if $X[1:|X|-1]$ generates a sequence shorter than $Y$ and
$X$ generates a sequence with $Y$ as a prefix (including $Y$ itself). Namely, let $f$ be the function of the random-stream algorithm,
them
$$|f(X[1:|X|-1])|<|Y|,$$
$$f(X)=Y\Delta \textrm{ with } \Delta\in\{0,1\}^*.$$
 To prove that the algorithm can generate
random-bit streams, we show that for any distinct binary sequences $Y_1, Y_2\in \{0,1\}^k$, the elements in $S_{Y_1}$ and those in $S_{Y_2}$ are
one-to-one mapping.

\begin{Lemma}\label{Stream_lemma1_3}
Let $f$ be the function of the random-stream algorithm. For any distinct binary sequences $Y_1, Y_2\in \{0,1\}^k$, if $X_A\in S_{Y_1}$, there are exactly one sequence $X_B\in S_{Y_2}$ such that
\begin{itemize}
  \item $X_B\equiv X_A$.
  \item $f(X_A)=Y_1\Delta$ and $f(X_B)=Y_2\Delta$ for some binary sequence $\Delta\in \{0,1\}^*$.
\end{itemize}
\end{Lemma}

\proof Let us prove this conclusion by induction. Here, we use $X_A'$ to denote the prefix of $X_A$ of length $|X_A|-1$ and use $\beta$ to denote
the last symbol of $X_A$. So $X_A=X_A'\beta$.

When $k=1$, if $X_A\in S_0$, we can write $f(X_A)$ as $0\Delta$ for some $\Delta \in \{0,1\}^*$. In this case, we assume that the status tree of $X_A'$ is $\mathcal{T}_A'$, and in which node $u$ generates the first bit $0$
when reading the symbol $\beta$. If we flip the label of $u$ from $0$ to $1$, we get another status tree, denoted by $\mathcal{T}_B'$. Using the same argument as Lemma \ref{Stream_lemma1_1}, we are able to construct a sequence $X_B'$ such that its status tree is $\mathcal{T}_B'$ and it does not generate any bits.
Concatenating $X_B'$ with $\beta$ results in a new sequence $X_B$, i.e., $X_B=X_B'\beta$, such that $X_B\equiv X_A$ and $f(X_B)=1\Delta$.
Similarly, for any sequence $X_B$ that yields $1$, i.e., $X_B\in S_1$, if $f(X_B)=1\Delta$, we can find a sequence $X_A\in S_0$ such that $X_A\equiv X_B$ and $f(X_A)=0\Delta$.
So we can say that the elements in $S_0$ and $S_1$ are one-to-one mapping.

Now we assume that all the elements in $S_{Y_1}$ and $S_{Y_2}$ are one-to-one mapping for all $Y_1, Y_2\in \{0,1\}^k$, then we show that this conclusion also holds for any $Y_1, Y_2\in \{0,1\}^{k+1}$. Two cases need to be considered.

1) $Y_1, Y_2$ end with the same bit. Without loss of generality, we assume this bit is $0$, then we can write $Y_1=Y_1'0$ and $Y_2=Y_2'0$.
If $X_A \in S_{Y_{1}'}$, then we can write $f(X_A)=Y_1'\Delta'$ in which the first bit of $\Delta'$ is $0$. According to
our assumption, there exists a sequence $X_B\in S_{Y_2'}$ such that $X_B\equiv X_A$ and $f(X_B)=Y_2'\Delta'$. In this case,
if we write $f(X_A)=Y_1\Delta=Y_1'0\Delta$, then $f(X_B)=Y_2'\Delta'=Y_2'0\Delta=Y_2\Delta$. So such a sequence $X_B$ satisfies our requirements.

If $X_A\notin S_{Y_{1}'}$, that means $Y_1'$ has been generated before reading the symbol $\beta$. Let us consider a prefix of $X_A$,
denote by $\overline{X_A}$, such that it yields $Y_1'$. In this case, $f(X_A')=Y_1'$ and we can write $X_A=\overline{X_A}Z$.
According to our assumption, there exists exactly one sequence $\overline{X_B}$ such that $\overline{X_B}\equiv\overline{X_A}$ and
$f(X_B')=Y_2'$. Since $\overline{X_A}$ and $\overline{X_B}$ induce
the same status tree, if we construct a sequence  $X_B=\overline{X_B}Z$, then $X_B\equiv X_A$ and $X_B$ generates
the same bits as $X_A$ when reading symbols from $Z$. It is easy to see that such a sequence $X_B$ satisfies our requirements.

Since this result is also true for the inverse case, if $Y_1, Y_2$ end with same bit the elements in $S_{Y_1}$ and $S_{Y_2}$ are one-to-one mapping.

2) Let us consider the case that $Y_1, Y_2$ end with different bits. Without loss of generality, we assume that $Y_1=Y_1'0$ and $Y_2=Y_2'1$.
According to the argument above, the elements in $S_{00...00}$ and $S_{Y_1'0}$ are one-to-one mapping; and the elements in $S_{00..01}$
and $S_{Y_2'1}$ are one-to-one mapping. So our task is to prove that the elements in $S_{00..00}$ and $S_{00...01}$ are one-to-one mapping.
For any sequence $X_A\in S_{00...00}$, let $X_A'$ be its prefix of length $|X_A|-1$.  Here, $X_A'$ generates only zeros whose length is at most $k$.
Let $\mathcal{T}_A'$ denote the status tree of $X_A'$ and let $u$ be the node in $\mathcal{T}_A'$ that generates the $k+1 th$ bit (zero) when reading the symbol $\beta$.
Then we can construct a new sequence $X_B'$ with status tree $\mathcal{T}_B'$ such that
\begin{itemize}
  \item $\mathcal{T}_B'$ and $\mathcal{T}_A'$ are the same except
  the label of $u$ is $0$ and the label of the node at the same position in $\mathcal{T}_B'$ is $1$.
  \item For each node $u$ in $\mathcal{T}_A'$, let $v$ be its corresponding node at the same position in $\mathcal{T}_B'$, then $u$ and $v$ generate the same bits.
\end{itemize}
The construction of $X_B'$ follows the proof of Lemma \ref{Stream_lemma1_2}. If we construct a sequence $X_B=X_B'\beta$,
it is not hard to show that $X_B$ satisfies our requirements, i.e.,
\begin{itemize}
  \item $X_B\equiv X_A$;
  \item $X_B'$ generates less than $k+1$ bits, i.e., $|f(X_B')|\leq k$;
  \item If $f(X_A)=0^k0\Delta$, then $f(X_B)=0^{k}1\Delta$, where $0^k$ is for $k$ zeros.
\end{itemize}

Also based on the inverse argument, we see that the elements in $S_{00..00}$ and $S_{00...01}$ are one-to-one mapping. So if $Y_1, Y_2$
end with different bits, the elements in $S_{Y_1}$ and $S_{Y_2}$ are one-to-one mapping.

Finally, we can conclude that
the elements in $S_{Y_1}$ and $S_{Y_2}$ are one-to-one mapping for any $Y_1, Y_2\in \{0,1\}^k$ with $k>0$.

This completes the proof.\hfill\QED

\vspace{0.05in}
\hspace{-0.15in}\textbf{Theorem \ref{Stream_theorem1}.} \emph{Given a source of biased coin with unknown probability, the random-stream algorithm generates a stream of random bits, i.e.,
for any $k>0$, if we stop running the algorithm after generating $k$ bits then these $k$ bits are independent and unbiased.}
\vspace{0.05in}

\proof
According to Lemma \ref{Stream_lemma1_3}, for any $Y_1, Y_2\in \{0,1\}^k$, the elements in $S_{Y_1}$ and $S_{Y_2}$ are one-to-one mapping.
If two sequences are one-to-one mapping, they have to be equivalent, which implies that their probabilities of being generated are the same.
Hence, the probability of generating a sequence in $S_{Y_1}$ is equal to that of generating a sequence in $S_{Y_2}$. It implies that
$Y_1$ and $Y_2$ have the same probability of being generated for a fixed number $k$. Since this is true for any $Y_1, Y_2\in \{0,1\}^k$,
the probability of generating an arbitrary binary sequence $Y\in \{0,1\}^k$ is $2^{-k}$. Finally, we have the statement in the theorem.

This completes the proof.\hfill\QED

\subsection{Proof of Theorem \ref{Stream_theorem2}}
\label{section_proof2}

\begin{Lemma}\label{stream_lemma2_1}
Given a stream of biased coin tosses, where the probability of generating H is $p$,
we run the random-stream algorithm until the number of coin tosses reaches $l$. In this case,
let $m$ be the number of random bits generated, then for any $\epsilon, \delta>0$, if $l$ is large enough, we have
that $$P[\frac{m-lH(p)}{lH(p)}<-\epsilon]<\delta,$$
where $$H(p)=-p\log_2 p -(1-p)\log_2(1-p)$$
is the entropy of the biased coin.
\end{Lemma}

\proof If we consider the case of fixed input length, then the random-stream algorithm is as efficient as Peres's algorithm asymptotically.
Using the same proof given in \cite{Peres1992} for Peres's algorithm,  we can get
$$\lim_{l\rightarrow\infty } \frac{E[m]}{l}=H(p).$$

Given a sequence of coin tosses of length $l$, we want to prove that for any $\epsilon>0$,
$$\lim_{l\rightarrow\infty}P[\frac{m-E[m]}{E[m]}<-\epsilon]=0.$$

To prove this result, we assume that this limitation holds for $\epsilon=\epsilon_1$, i.e., for any $\delta>0$, if $l$ is large enough, then
$$P[\frac{m-E[m]}{E[m]}<-\epsilon_1]<\delta.$$
Under this assumption, we show that there always exists $\epsilon_2<\epsilon_1$ such that the limitation also holds for $\epsilon=\epsilon_2$.
Hence, the value of $\epsilon$ can be arbitrarily small.

In the random-stream algorithm, $l$ is the number of symbols (coin tosses) received by the root. Let $m_1$ be the number of random bits generated by
the root, $m_{(l)}$ be the number of random bits generated by its left subtree and $m_{(r)}$ be the number of random bits
generated by its right subtree. Then it is easy to get
$$m=m_1+m_{(l)}+m_{(r)}.$$

Since the $m_1$ random bits generated by the root node are independent, we can always make $l$ large enough such that
$$P[\frac{m_1-E[m_1]}{E[m_1]}<-\epsilon_1/2]<\delta/3.$$
At the same time, by making $l$ large enough, we can make both $m_{(l)}$ and $m_{(r)}$ large enough such that (based on our assumption)
$$P[\frac{m_{(l)}-E[m_{(l)}]}{E[m_{(l)}]}<-\epsilon_1]<\delta/3$$
and
$$P[\frac{m_{(r)}-E[m_{(r)}]}{E[m_{(r)}]}<-\epsilon_1]<\delta/3.$$

Based on the three inequalities above, we can get
$$P[m-E[m]\leq -\epsilon_1 (\frac{E[m_1]}{2} +E[m_{(l)}]+E[m_{(r)}])]<\delta.$$

If we set
$$\epsilon_2=\epsilon_1\frac{\frac{E[m_1]}{2} +E[m_{(l)}]+E[m_{(r)}]}{E[m_1+m_{(l)}+m_{(r)}]},$$
then
$$P[\frac{m-E[m]}{E[m]}<-\epsilon_2]<\delta.$$

Given the probability $p$ of the coin, when $l$ is large,
 $$E[m_1]=\Theta(E[m]), E[m_{(l)}]=\Theta(E[m]), E[m_{(r)}]=\Theta(E[m]),$$
which implies that $\epsilon_2<\epsilon_1$.

So we can conclude that for any $\epsilon>0, \delta>0$, if $l$ is large enough then
$$P[\frac{m-E[m]}{E[m]}<-\epsilon]<\delta.$$
And based on the fact that $E[m]\rightarrow lH(p)$, we get the result in the lemma.
\hfill\QED

\vspace{0.05in}
\hspace{-0.15in}\textbf{Theorem \ref{Stream_theorem2}.} \emph{Given a biased coin with probability $p$ being H, let $n$ be the number of coin tosses required for generating $k$ random bits in the random-stream algorithm, then
$$\lim_{k\rightarrow\infty}\frac{E[n]}{k}= \frac{1}{H(p)}.$$}

\proof For any $\epsilon, \delta>0$, we set $l=\frac{k}{H(p)}(1+\epsilon)$, according to
the conclusion of the previous lemma, with probability at least $1-\delta$, the output length is at least $k$
if the input length $l$ is fixed and large enough. In another word, if the output length is $k$, which is fixed, then
with probability at least $1-\delta$, the input length $n\leq l$.

So with probability less than $\delta$, we require more than $l$ coin tosses. The worst case is that we did not generate any bits for the first $l$ coin tosses.  In this case,
we need to generate $k$ more random bits. As a result, the expected number of coin tosses required is at most $l+E[n]$.

Based on the analysis above, we derive
$$E[n]\leq (1-\delta)l + (\delta) (l+E[n]),$$
then
$$E[n]\leq \frac{l}{1-\delta}=\frac{k}{H(p)}\frac{(1+\epsilon)}{(1-\delta)}.$$

Since $\epsilon,\delta$ can be arbitrarily small when $l$ (or $k$) is large enough
$$\lim_{k\rightarrow\infty}\frac{E[n]}{k}\leq \frac{1}{H(p)}.$$

Based on Shannon's theory \cite{Cover2006}, it is impossible to generate $k$ random bits from a source with expected entropy less than $k$.
Hence
$$\lim_{k\rightarrow\infty}E[nH(p)] \geq k,$$
i.e.,
$$\lim_{k\rightarrow\infty}\frac{E[n]}{k}\geq \frac{1}{H(p)}.$$

Finally, we get the conclusion in the theorem. This completes the proof.
\hfill\QED

\subsection{Proof of Theorem \ref{Stream_theorem7}}
\label{section_proof3}

The proof of Theorem \ref{Stream_theorem7} is very similar as the proof of Theorem \ref{Stream_theorem1}. Let $S_Y$ with $Y\in \{0,1\}^k$ denote the
set consisting of all the binary sequences yielding $Y$ in the random-stream algorithm with limited maximum depth. Then for any distinct binary sequences $Y_1, Y_2\in \{0,1\}^k$, the elements
in $S_{Y_1}$ and those in $S_{Y_2}$ are one-to-one mapping. Specifically, we can get the following lemma.

\begin{Lemma} Let $f$ be the function of the random-stream algorithm with maximum depth $d$. For any distinct binary sequences $Y_1, Y_2\in \{0,1\}^k$,
if $X_A\in S_{Y_1}$, there exists one sequence $X_B\in S_{Y_2}$ such that
\begin{itemize}
  \item $X_A\equiv X_B$.
  \item Let $\mathcal{T}_A$ be the status tree of $X_A$ and $\mathcal{T}_B$ be the status tree of $X_B$. For any node $u$ with depth larger than $d$ in $\mathcal{T}_A$,
  let $v$ be its corresponding node in $\mathcal{T}_B$ at the same position, then $u$ and $v$ generate the same bits.
  \item $f(X_A)=Y_1\Delta$ and $f(X_B)=Y_2\Delta$ for some binary sequence $\Delta\in \{0,1\}^*$.
\end{itemize}
\end{Lemma}

\proof The proof of this lemma is a simple modification of that for Lemma \ref{Stream_lemma1_3}, which is by induction. A simple sketch is given as follows.

First, similar as the proof for Lemma \ref{Stream_lemma1_3}, it can be proved that: when $k=1$, for any sequence $X_A \in S_0$,
there exists one sequence $X_B\in S_1$ such that $X_A, X_B$ satisfy the conditions in the lemma, and vice versa. So we can
say that the elements in $S_0$ and $S_1$ are one-to-one mapping. Then we assume that all the elements in $S_{Y_1}$ and $S_{Y_2}$ are one-to-one mapping for all $Y_1, Y_2 \in \{0, 1\}^k$,
then we show that this conclusion also holds for any $Y_1, Y_2 \in\{0, 1\}^{k+1}$. Two cases need to be considered.

1) $Y_1, Y_2$ end with the same bit. Without loss of generality, we assume this bit is $0$, then we can
write $Y_1=Y_1'0$ and $Y_2=Y_2'0$.

If $X_A\in S_{Y_1'}$, then according to our assumption, it is easy to prove the conclusion, i.e., there exists a sequence $X_B$ satisfies the conditions.

If $X_A\notin S_{Y_1'}$, then we can write $X_A=\overline{X_A}Z$ and $\overline{X_A}\in S_{Y_1'}$. According to our assumption,
for the sequence $\overline{X_A}$, we can find its mapping $\overline{X_B}$ such that (1) $\overline{X_A}\equiv \overline{X_B}$; (2)
$\overline{X_A},\overline{X_B}$ induce the same status tree and their corresponding nodes with depth larger than $d$ generate the same bits;
and (3) $f(\overline{X_A})=Y_1'$ and $f(\overline{X_B})=Y_2'$. If we construct a sequence $X_B = X_BZ$, it will satisfy all the conditions in the lemma.

Since this result is also true for the inverse case, if $Y_1, Y_2$ end with same bit, the elements in $S_{Y_1}$
and $S_{Y_2}$ are one-to-one mapping.

2) $Y_1,Y_2$ end with different bits. Without loss of generality, we assume
that $Y_1 =Y_1'0$ and $Y_2=Y_2'1$. According to the argument above, the elements in
$S_{0^{k}0}$ and $S_{Y_1}$ are one-to-one mapping; and the elements in $S_{0^{k}1}$ and
$S_{Y_2}$ are one-to-one mapping. So we only need to prove that the elements in $S_{0^{k}0}$ and
$S_{0^{k}1}$ are one-to-one mapping. In this case, for any $X_A\in S_{0^{k-1}0}$, let $X_A=X_A'\beta$ with a single symbol $\beta$.
Then  $X_A'$ generates only zeros whose length is at most $k$. Let $\mathcal{T}_A'$ denote the status tree of $X_A'$ and let $u$ be the node in $\mathcal{T}_A'$ that generates the $k +1$th bit (zero) when reading the symbol $\beta$. Note that the depth of $u$ is at most $d$. In this case, we can construct a new sequence $X_B'$ with status tree $\mathcal{T}_B'$ such that
\begin{itemize}
  \item $\mathcal{T}_B'$ and $\mathcal{T}_A'$ are the same except
  the label of $u$ is $0$ and the label of the node at the same position in $\mathcal{T}_B'$ is $1$.
  \item For each node $u$ in $\mathcal{T}_A'$, let $v$ be its corresponding node at the same position in $\mathcal{T}_B'$, then $u$ and $v$ generate the same bits.
\end{itemize}
Then we can prove that the sequence $X_B=X_B'\beta$ satisfies our all our conditions in the lemma. Also based on the inverse argument, we can claim that the elements in $S_{0^k0}$ and $S_{0^k1}$ are one-to-one
mapping.

Finally, we can conclude that the elements in $S_{Y_1}$ and $S_{Y_2}$ are one-to-one mapping for any $Y_1, Y_2 \in\{0, 1\}^k$ with $k > 0$.

This completes the proof.\hfill\QED\vspace{0.05in}

From the above lemma, it is easy to get Theorem \ref{Stream_theorem7}.

\vspace{0.05in}
\hspace{-0.15in}\textbf{Theorem \ref{Stream_theorem7}.} \emph{Given a source of biased coin with unknown probability, the random-stream algorithm with maximum depth $d$ generates  a stream of random bits, i.e.,
for any $k>0$, if we stop running the algorithm after generating $k$ bits then these $k$ bits are independent and unbiased.}
\vspace{0.05in}

\proof We can apply the same procedure of proving Theorem \ref{Stream_theorem2}. \hfill\QED

\subsection{Proof of Theorem \ref{stream_theorem4}}

Similar to the proof of Theorem \ref{Stream_theorem2}, we first consider the case that the input length is fixed.

\begin{Lemma}\label{stream_lemma2_2}
Given a stream of biased coin tosses, where the probability of generating H is $p$,
we run the random-stream algorithm with maximum depth $d$ until the number of coin tosses reaches $l$. In this case,
let $m$ be the number of random bits generated, then for any $\epsilon, \delta>0$, if $l$ is large enough, we have
that $$P[\frac{m-l\rho_d(p)}{l\rho_d(p)}<-\epsilon]<\delta,$$
where $\rho_d(p)$ is given in (\ref{stream_equ_efficiencydepthd}).
\end{Lemma}

\proof
Let $\rho_d(p)$ be the asymptotic expected number of random bits generated per coin toss when the random-stream algorithm has maximum depth $d$ and
the probability of the biased coin is $p$. Then
$$\lim_{l\rightarrow\infty}\frac{E[m]}{l}=\rho_d(p).$$
When the fixed input length $l$ is large enough, the random-stream algorithm generates approximately
$l\rho_d(p)$ random bits, which are generated by the root node, the left subtree (subtree rooted at root's left child) and the right subtree (subtree rooted at the root's right child). Considering the root node, it generates approximately $lpq$ random bits with $q=1-p$. Meanwhile, the root node passes approximately $\frac{l}{2}$ messages (H or T) to its left child, where the messages are independent and the probability of H is $p^2+q^2$;
and the root node passes approximately $\frac{l}{2}(p^2+q^2)$ messages (H or T) to its right child, where the messages are independent and
the probability of H is $\frac{p^2}{p^2+q^2}$. As a result, according to the definition of $\rho_d$, the left
subtree generates approximately $\frac{l}{2} \rho_{d-1}(p^2+q^2)$ random bits, and the right subtree generates
approximately $\frac{l}{2}(p^2+q^2) \rho_{d-1}(\frac{p^2}{p^2+q^2})$ random bits. As $l\rightarrow\infty$, we have
$$\lim_{l\rightarrow\infty}\frac{l\rho_d(p)}{lpq+\frac{l}{2} \rho_{d-1}(p^2+q^2)+\frac{l}{2}(p^2+q^2) \rho_{d-1}(\frac{p^2}{p^2+q^2})}=1.$$
It yields
$$\rho_d(p)=pq+\frac{1}{2} \rho_{d-1}(p^2+q^2)+\frac{1}{2}(p^2+q^2) \rho_{d-1}(\frac{p^2}{p^2+q^2}).$$
So we can calculate $\rho_d(p)$ by iteration. When $d=0$, the status tree has the single root node, and it is easy to get $\rho_0(p)=pq$.

Then, following the proof of Lemma \ref{stream_lemma2_1}, for any $\epsilon, \delta>0$, if $l$ is large enough, we have
that $$P[\frac{m-E[m]}{E[m]}<-\epsilon]<\delta.$$

So we can get the conclusion in the lemma. This completes the proof.
\hfill\QED\vspace{0.05in}

From the above lemma, we can get Theorem  \ref{stream_theorem4}, that is,\vspace{0.05in}

\hspace{-0.15in}\textbf{Theorem \ref{stream_theorem4}.} \emph{ When the maximum depth of the tree is $d$ and the probability of the biased coin is $p$ of being H, the expected number of coin tosses
required per random bit is
$$\frac{1}{\rho_d(p)}$$ asymptotically, where $\rho_d(p)$ can be obtained by iterating
\begin{equation*}\rho_d(p)=pq+\frac{1}{2}\rho_{d-1}(p^2+q^2)+\frac{1}{2}(p^2+q^2)\rho_{d-1}(\frac{p^2}{p^2+q^2})\quad(\ref{stream_equ_efficiencydepthd})\end{equation*}
with $q=1-p$ and $\rho_0(p)=pq$.}\vspace{0.05in}

\proof We can apply the same procedure of proving Theorem \ref{Stream_theorem1} except we apply Lemma  \ref{stream_lemma2_2} instead of Lemma \ref{stream_lemma2_1}.\hfill\QED

\section{Generalized Random-Stream Algorithm}
\label{Stream_section_generalization}

\subsection{Preliminary}

In \cite{Zhou2012LoadedDice}, we introduced a universal scheme for transforming an arbitrary algorithm for generating random bits from
a sequence of biased coin tosses to manage the general source of an $m$-sided die.
This scheme works when the input is a sequence of fixed length; in this section, we study how to modify this scheme to generate
random-bit streams from $m$-sided dice. For sake of completeness we describe the original scheme here.

\begin{figure}[!t]
\centering
\includegraphics[width=2.4in]{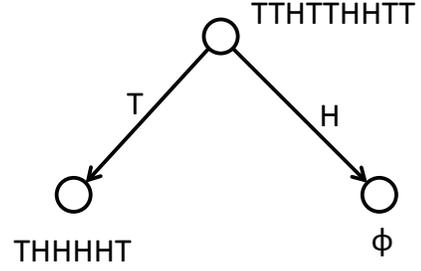}
\caption{An instance of binarization tree.}
\label{fig_prefixtree}
\end{figure}

The main idea of the scheme is to convert
a sequence with alphabet larger than two, written as
$$X=x_1x_2...x_n\in \{0,1,...,m-1\}^n,$$ into multiple binary sequences.  To do this, we create a binary tree, called a binarization tree, in which
each node is labeled with a binary sequence of H and T.
Given the binary representations of $x_i$ for all $1\leq i\leq n$,
the path of each node in the tree indicates a prefix, and the binary sequence labeled at this node
consists of all the bits (H or T) following the prefix in the binary representations of $x_1, x_2, ..., x_n$ (if it exists).
Fig.~\ref{fig_prefixtree} is an instance of binarization tree when the input sequence is $X=012112210$, produced by a
$3$-sided die. To see this, we write each symbol  (die roll) into a binary representation of length two,
hence $X$ can be represented as
$$\textrm{TT,TH,HT,TH,TH,HT,HT,TH,TT}.$$
Only collecting the first bits of all the symbols yields an independent binary sequence $$X_{\phi}=\textrm{TTHTTHHTT},$$ which is labeled on the root node; 
Collecting the second bits following T, we get another independent binary sequence $$X_\textrm{T}=\textrm{THHHHT},$$ which is labeled on the left child of the root node.

The universal scheme says that we can `treat' each binary sequence labeled on the binarization tree as a sequence of biased coin tosses: Let
$\Psi$ be any algorithm that can generate random bits from an arbitrary biased coin, then applying $\Psi$ to each of the sequences labeled on the binarization tree and concatenating their outputs together results in an  independent and unbiased sequence, namely, a sequence of random bits.

Specifically, given the number of sides $m$ of a loaded die, the depth of the binarization tree is $b=\lceil\log_2 m \rceil-1$.
Let $\Upsilon_b$ denote the set consisting of all the binary sequences of length at most $b$, i.e.,
$$\Upsilon_b=\{\textrm{$\phi$, T, H, TT, TH, HT, HH, ..., HHH...HH}\}.$$
Given $X\in\{0,1,...,m-1\}^n$, let $X_\gamma$ denote the binary sequence labeled on a node corresponding to a prefix $\gamma$ in the binarization tree, then
we get a group of binary sequences
$$X_\phi, X_\textrm{T}, X_\textrm{H}, X_\textrm{TT}, X_\textrm{TH}, X_\textrm{HT}, X_\textrm{HH}, ...$$
For any function $\Psi$ that generates random bits from a fixed number of coin tosses, we can generate random bits from $X$ by calculating
$$\Psi(X_\phi)+\Psi(X_\textrm{T})+\Psi(X_\textrm{H})+\Psi(X_\textrm{TT})+\Psi(X_\textrm{TH})+...,$$
where $A+B$ is the concatenation of $A$ and $B$.

So in the above example, the output of $X=012112210$ is $\Psi(X_\phi)+\Psi(X_\textrm{T})$, i.e.,
$$\Psi(\textrm{TTHTTHHTT})+\Psi(\textrm{THHHHT}).$$
This conclusion is simple, but not obvious, since the binary sequences labeled on the same binarization tree are correlated with each other.

\subsection{Generalized Random-Stream Algorithm}

We want to generalize the random-stream algorithm to generate random-bit streams from an $m$-sided die.
Using the similar idea as above, we convert the input stream into
multiple binary streams, where each binary stream corresponds to a node in the binalization tree. We apply the random-stream algorithm
to all these binary streams individually, and for each stream we create a status tree for storing state information.
When we read a dice roll of $m$ sides from the source, we pass all the $\log_2 m$ bits of its binary representation to $\lceil \log_2 m\rceil$ different streams that corresponds to
a path in the binalization tree. Then we process all these $\lceil\log_2 m\rceil$ streams from top to bottom along that path.
In this way, a single binary stream is produced. While each node in the binalization tree generates a random-bit stream,
the resulting single stream is a mixture of these random-bit streams. But it is not obvious whether
the resulting stream is a random-bit stream or not, since the values of the bits generated affect their orders.

The following example is constructed for demonstrating this algorithm.

\begin{figure}[!t]
\centering
\includegraphics[width=3.6in]{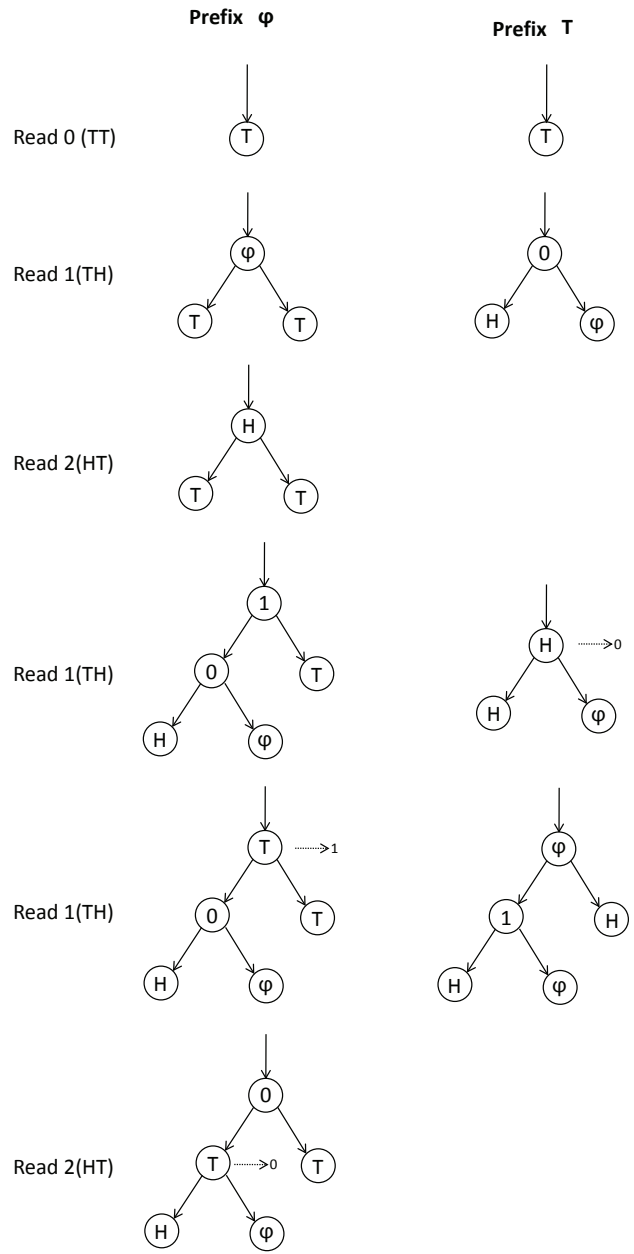}
\caption{The changes of status trees in the generalized random-stream algorithm when the input stream is $012112210...$.}
\label{fig_generalized}
\end{figure}

Let us consider a stream of symbols generated from a $3$-sided die,
$$012112210...$$
Instead of storing a binary sequence at each node in the binalization tree,
we associate each node with a status tree corresponding to a random-stream algorithm.
Here, we get two nontrivial binary streams
$$\textrm{TTHTTHHTT..., THHHHT...}$$
corresponding to prefix $\phi$ and $\textrm{T}$ respectively,
Fig. \ref{fig_generalized} demonstrates how the status trees change when we read symbols one by one.
For instance, when the $4$th symbol $1$(TH) is read, it passes T to the root node (corresponding to
the prefix $\phi$) and passes H to the left child of the root node (corresponding to the prefix T)
of the binalization tree. Based on the rules of the random-stream algorithm, we modify the status trees associated with these two nodes.
During this process, a bit $0$ is generated.

Finally, this scheme generates a stream of bits $010...$, where the first bit is generated after reading the $4$th
symbol, the second bit is generated after reading the $5$th symbol, ...
We call this scheme as the generalized random-stream algorithm. As we expected, this algorithm
can generate a stream of random bits from an arbitrary loaded die with $m\geq 2$ sides.

\begin{Theorem} \label{Stream_theorem5}
Given a loaded die with $m\geq 2$ sides, if we stop running the generalized random-stream algorithm after generating $k$ bits,
then these $k$ bits are independent and unbiased.
\end{Theorem}

The proof of the above theorem is given in Subsection \ref{Stream_section_provetheorem5}.

Since the random-stream algorithm is as efficient as Peres's algorithm asymptotically, we can prove that the generalized random-stream algorithm is also asymptotically optimal.

\begin{Theorem} \label{Stream_theorem6}
 Given an $m$-sided die with probability distribution $\rho=(p_0, p_1, ..., p_{m-1})$,
let $n$ be the number of symbols (dice rolls) used in the generalized random-stream algorithm and let $k$ be the number of
random bits generated, then
 $$\lim_{k\rightarrow\infty}\frac{E[n]}{k}=\frac{1}{H(p_0,p_1,...,p_{m-1})},$$
where $$H(p_0,p_1,...,p_{m-1})=\sum_{i=0}^{m-1}p_i\log_2\frac{1}{p_i}$$
is the entropy of the $m$-sided die.
\end{Theorem}

\proof First, according to Shannon's theory, it is easy to get that
$$\lim_{k\rightarrow\infty}\frac{E[n]}{k}\geq
\frac{1}{H(p_0,p_1,...,p_{m-1})}.$$

Now, we let $$n=\frac{k}{H(p_0,p_1,...,p_{m-1})}(1+\epsilon)$$ with an arbitrary $\epsilon>0$. Following the proof of Theorem 7 in \cite{Zhou2012LoadedDice},
it can be shown that when  $k$ is large enough,
the algorithm generates more than $k$ random bits with probability at least $1-\delta$ with any $\delta>0$.
Then using the same argument in Theorem \ref{Stream_theorem2}, we can get
$$\lim_{k\rightarrow\infty}\frac{E[n]}{k}\leq
\frac{1}{H(p_0,p_1,...,p_{m-1})}\frac{1+\epsilon}{1-\delta},$$
for any $\epsilon,\delta>0$.

Hence, we can get the conclusion in the theorem.
\hfill\QED

Of source, we can limit the depths of all the status trees for saving space, with proof emitted. It can be seen that given a loaded die of $m$ sides, the space usage is proportional to $m$ and the expected computational
time is proportional to $\log m$.

\subsection{Proof of Theorem \ref{Stream_theorem5}}
\label{Stream_section_provetheorem5}

Here, we want to prove that the generalized random-stream algorithm generates a stream of random bits from an arbitrary $m$-sided die. Similar as above, we let
$S_Y$ with $Y\in \{0,1\}^k$ denote the set consisting of all the sequences yielding $Y$. Here, we say that a sequence $X$
yields $Y$ if and only if $X[1:|X|-1]$ generates a sequence shorter than $Y$ and $X$ generates a sequence with $Y$ as a prefix (including $Y$ itself). We would like
to show that the elements in $S_{Y_1}$ and those in $S_{Y_2}$ are one-to-one mapping if $Y_1$ and $Y_2$ have
the same length.

\begin{Definition}\label{definition_2}
Two sequences $X_A, X_B\in \{0,1,...,m-1\}^*$ with $m>2$ are equivalent, denoted by $X_A\equiv X_B$, if and only ${X}_\gamma^{A}\equiv {X}_\gamma^B$ for all
$\gamma\in \Upsilon_b$, where ${X}_\gamma^{A}$ is the binary sequence labeled on a node corresponding to a prefix $\gamma$ in the binalization tree induced by $X_A$, and
the equivalence of ${X}_\gamma^{A}$ and ${X}_\gamma^{B}$ was given in Definition \ref{definition_1}.
\end{Definition}

\begin{Lemma} \emph{\cite{Zhou2012LoadedDice}} \label{Stream_lemma_biasedcoin}
Let $\{X_{\gamma}^A\}$ with $\gamma\in \Upsilon_b$ be the binary sequences labeled on the binarization tree of $X_A\in \{0,1,...,m-1\}^n$ as defined above.
Assume $X_\gamma^B$ is a permutation of $X_\gamma^A$ for all $\gamma\in \Upsilon_b$, then there exists exactly one sequence $X_B\in \{0,1,...,m-1\}^n$ such that it yields a binarization tree that labels $\{X_{\gamma}^B\}$ with $\gamma\in \Upsilon_b$.
\end{Lemma}

\proof The proof is provided in \cite{Zhou2012LoadedDice}. \hfill\QED

\begin{Lemma} Let $f$ be the function of the generalized random-stream algorithm, and let $X_A$ be a sequence produced by an $m$-sided die. For any distinct sequences $Y_1, Y_2\in \{0,1\}^k$, if
$X_A \in S_{Y_1}$, there are exactly one sequence $X_B\in S_{Y_2}$ such that
\begin{itemize}
  \item $X_B\equiv X_A$.
  \item $f(X_A)=Y_1\Delta$ and $f(X_B)=Y_2\Delta$ for some binary sequence $\Delta$.
\end{itemize}
\end{Lemma}

\proof The idea of the proof is to combine the proof of Lemma \ref{Stream_lemma1_3} with the
result in Lemma \ref{Stream_lemma_biasedcoin}.

Let us prove this conclusion by induction.
Here, we use $X_A'$ to denote
the prefix of $X_A$ of length $|X_A|-1$ and use $\beta$ to denote the last symbol of $X_A$. So $X_A=X_A'\beta$.
${{X}_\gamma^{A}}$ is the binary sequence labeled on a node corresponding to a prefix $\gamma$ in the binalization tree induced by $X_A'$,
and the status tree of ${{X}_\gamma^{A}}'$ with $\gamma \in \Upsilon_b$ is denoted as ${\mathcal{T}_\gamma^{A}}$.

When $k=1$, if $X_A\in S_0$, we can write $f(X_A)$ as $0\Delta$. In this case, let $u$ in ${\mathcal{T}_\theta^{A}}$ with $\theta\in \Upsilon_b$ be the node that generates the first bit $0$. If
we flip the label of $u$ from $0$ to $1$, we get another status tree $\mathcal{T}_\theta^B$. Using the same argument in Lemma \ref{Stream_lemma1_2},
we are able to construct a sequence $X_\theta^B$ such that its status tree is $\mathcal{T}_\theta^B$ and it does not generate any bits. Here,
$X_\theta^B$ is a permutation of $X_\theta^A$.
From $X_{\phi}^A, X_{T}^A, ..., X_{\theta}^B,...$, we can construct a sequence $X_B'$ uniquely following the proof of Lemma \ref{Stream_lemma_biasedcoin} (see \cite{Zhou2012LoadedDice}).
Concatenating $X_B'$ with $\beta$ results in a new sequence $X_B$, i.e., $X_B=X_B'\beta$ such that $X_B\equiv X_A$ and $f(X_B)=1\Delta$.
Inversely, we can get the same result. It shows that the elements in $S_0$ and $S_1$ are one-to-one mapping.

Now we assume that the conclusion holds for all $Y_1, Y_2\in \{0,1\}^k$, then we show that it also holds for any $Y_1, Y_2\in \{0,1\}^{k+1}$. Two cases need to be considered.

1) $Y_1, Y_2$ end with the same bit. Without loss of generality, we assume that this bit is $0$, then we can write $Y_1=Y_1'0$ and $Y_2=Y_2'0$.
If $X_A$ yields $Y_1$, based on our assumption, it is easy to see that there exists a sequence $X_B$ satisfies our requirements.
If $X_A$ does not yield $Y_1$, that means $Y_1'$ has been generated before reading the symbol $\beta$. Let us consider a prefix of $X_A$, denote by
$\overline{X_A}$, such that it yields $Y_1'$. In this case, $f(X_A')=Y_1'$ and we can write $X_A=X_A'Z$. According to our assumption,
there exists exactly one sequence $\overline{X_B}$ such that $\overline{X_B}\equiv\overline{X_A}$ and
$f(X_B')=Y_2'$. Since $\overline{X_A}$ and $\overline{X_B}$ lead to
the same binalization tree (all the status trees at the same positions are the same), if we construct a sequence  $X_B=\overline{X_B}Z$, then $X_B\equiv X_A$ and $X_B$ generates the same bits as $X_A$ when reading symbols from $Z$. It is easy to see that such a sequence $X_B$ satisfies our requirements.

Since this result is also true for the inverse case, if $Y_1, Y_2\in \{0,1\}^{k+1}$ end with the same bit, the elements in $S_{Y_1}$ and $S_{Y_2}$ are one-to-one mapping.

2) Let us consider the case that $Y_1, Y_2$ end with different bits. Without loss of generality, we assume that $Y_1=Y_1'0$ and $Y_2=Y_2'1$.
According to the argument above, the elements in $S_{00...00}$ and $S_{Y_1'0}$ are one-to-one mapping; the elements in $S_{00..01}$
and $S_{Y_2'1}$ are one-to-one mapping. So our task is to prove that the elements in $S_{00..00}$ and $S_{00...01}$ are one-to-one mapping.
For any sequence $X_A\in S_{00...00}$, let $X_A'$ be its prefix of length $|X_A|-1$.  Here,
$X_A'$ generates only zeros whose length is at most $k$.
Let $\mathcal{T}_\theta^A$ denote one of the status trees such that $u\in T_\theta^A$ is the node that generates that $k+1 th$ bit (zero) when  reading the symbol $\beta$.
Then we can construct a new sequence $X_B'$ such that
\begin{itemize}
 \item Let $\{X_\gamma^B\}$ with $\gamma \in \Upsilon_b$ be the binary sequences induced by $X_B'$, and let
  $\mathcal{T}_\gamma^B$ be the status tree of $X_\gamma^B$. The binalization trees of $X_A'$ and $X_B'$ are the same (all the status trees at the same positions are the same), except
  the label of $u$ is $0$ and the label of its corresponding node $v$ in $\mathcal{T}_\theta^B$ is $1$.
  \item Each node $u$ in $\mathcal{T}_\gamma^B$ generates
  the same bits as its corresponding node $v$ in $\mathcal{T}_\gamma^A$ for all $\gamma \in \Upsilon_b$.
\end{itemize}
The construction of $X_B'$ follows the proof of Lemma \ref{Stream_lemma1_1} and then Lemma \ref{Stream_lemma_biasedcoin}. If we construct a sequence $X_B=X_B'\beta$,
it is not hard to show that $X_B$ satisfies our requirements, i.e.,
\begin{itemize}
  \item $X_B\equiv X_A$;
  \item $X_B'$ generates less than $k+1$ bits, i.e., $|f(X_B')|\leq k$;
  \item If $f(X_A)=Y_1\Delta=Y_1'0\Delta$, then $f(X_B)=Y_2'1\Delta=Y_2\Delta$.
\end{itemize}

Also based on the inverse argument, we see that the elements in $S_{00..00}$ and $S_{00...01}$ are one-to-one mapping.

Finally, we can conclude that
the elements in $S_{Y_1}$ and $S_{Y_2}$ are one-to-one mapping for any $Y_1, Y_2\in \{0,1\}^k$ with $k>0$.

This completes the proof.\hfill\QED\vspace{0.05in}

Based on the above result and the argument for Theorem \ref{Stream_theorem1}, we can easily prove Theorem \ref{Stream_theorem5}.

\vspace{0.05in}
\hspace{-0.15in}\textbf{Theorem \ref{Stream_theorem5}.} \emph{Given a loaded die with $m\geq 2$ sides, if we stop running the generalized random-stream algorithm after generating $k$ bits,
then these $k$ bits are independent and unbiased.}

\section{Extension for Markov Chains}
\label{Stream_section_Markov}

In this section, we study how to efficiently generate random-bit streams from Markov chains. The nonstream case  was studied by Samuelson \cite{Samuelson1968}, Blum \cite{Blum1986} and later generalized
by Zhou and Bruck \cite{Zhou2012Markov}. Here, using the techniques developed in \cite{Zhou2012Markov}, and applying the techniques introduced in this paper, we are able to generate random-bit streams from Markov chains. We present the algorithm briefly.

For a given Markov chain, it generates a stream of states, denoted by $x_1x_2x_3...\in \{s_1,s_2,...,s_m\}^*$. We can treat each state, say $s$, as a die and consider the `next states' (the states the chain has transitioned to after being at state $s$) as the results of a die roll, called the exit of $s$. For all $s\in \{s_1,s_2,...,s_m\}$, if we only consider the exits of $s$, they form a stream. So we have total $m$ streams corresponding to the exits of $s_1, s_2, ..., s_m$ respectively.  For example, assume the input is
$$X=s_1 s_4 s_2 s_1 s_3 s_2 s_3 s_1 s_1 s_2 s_3 s_4 s_1 ...$$

If we consider the states following $s_1$, we get a stream as the set of states in boldface:
$$X = s_1 \textbf{s}_\textbf{{4}} s_2 s_1 \textbf{s}_\textbf{{3}} s_2 s_3 s_1 \textbf{s}_\textbf{{1}} \textbf{s}_\textbf{{2}} s_3 s_4 s_1...$$

Hence the four streams are
$$s_4 s_3 s_1s_2 ... , s_1 s_3 s_3..., s_2s_1 s_4..., s_2 s_1...$$

The generalized random-stream algorithm is applied to each stream separately for generating random-bit streams.
Here, when we get an exit of a state $s$, we should not directly pass it to the generalized random-stream algorithm that corresponds to the state $s$.
Instead, it waits until we get the next exit of the state $s$. In another word, we keep the current exit in pending. In the above example,
after we read $s_1 s_4 s_2 s_1 s_3 s_2 s_3 s_1 s_1 s_2 s_3 s_4 s_1$, $s_4s_3s_1$ has been passed to the generalized random-stream algorithm
corresponding to $s_1$, $s_1s_3$ has been passed to the generalized random-stream algorithm corresponding to $s_2$,...the most recent exit of each state, namely
$s_2, s_3, s_4, s_1$ are in pending.
Finally, we mix all the bits generated from different streams based on their natural generating order. As a result, we get a stream of random bits from an arbitrary Markov chain, and it achieves
the information-theoretic upper bound on efficiency.

Now, we call this algorithm the random-stream algorithm for Markov chains, and we describe it as follows.

\vspace{0.05in}
\begin{list}{\labelitemi}{\leftmargin=0.5em}
\renewcommand{\labelitemi}{}
  \item \textbf{Input:} A stream $X=x_1x_2x_3...$ produced by a Markov chain, where $x_i\in S=\{s_1,s_2,...,s_m\}$.
  \item \textbf{Output:} A stream of $0'$s and $1'$s.
  \item \textbf{Main Function:}
  \begin{algorithmic}
\STATE Let $\Phi_i$ be the generalized random-stream algorithm for the exits of $s_i$ for $1\leq i\leq m$, and
$\theta_i$ be the pending exit of $s_i$ for $1\leq i\leq m$.
\STATE Set $\theta_i=\phi$ for $1\leq i\leq m$.
\FOR{ each symbol $x_j$ read from the Markov chain }
    \IF{$x_{j-1}=s_i$}
        \IF{$\theta_i\neq \phi$}
        \STATE Input $\theta_i$ to $\Phi_i$ for processing.
        \ENDIF
        \STATE Set $\theta_i=x_j$.
    \ENDIF
\ENDFOR
\end{algorithmic}
\end{list}

\begin{Theorem} Given a source of a Markov chain with unknown transition probabilities,
the random-stream algorithm for Markov chains generates a stream of random bits, i.e.,
for any $k>0$, if we stop running the algorithm after generating $k$ bits then these $k$ bits are independent and unbiased.
\end{Theorem}

The proof of the above theorem is a simple extension of the proof for Theorem \ref{Stream_theorem5}. Let $S_Y$ denote the set of input sequences that yield a binary sequence $Y$. Our main idea is still to prove that all the elements in $S_{Y_1}$ and $S_{Y_2}$ are one-to-one mapping for all $Y_1, Y_2\in \{0,1\}^k$ with $k>0$. The detailed proof is a little complex, but it is not difficult; we only need to follow the proof of Theorem \ref{Stream_theorem5} and combine it with the following result from \cite{Zhou2012Markov}.  Here, we omit the detailed proof.

\begin{Lemma} Given an input sequence $X=x_1x_2...x_N\in \{s_1,s_2,...,s_m\}^N$ that produced from a Markov chain, let
$\pi_i(X)$ be the exit sequence of $s_i$ (the symbols following $s_i$) for $1\leq i\leq m$. Assume that  $[\Lambda_1,\Lambda_2,...,\Lambda_n]$
is an arbitrary collection of exit sequences such that $\Lambda_i$ and $\pi_i(X)$ are permutations and
they have the same last element for all $1\leq i\leq m$. Then there exists a sequence $X'=x_1'x_2'...x_N'\in \{s_1,s_2,...,s_m\}^N$ such that
$x_1'=x_1$ and $\pi_i(X')=\Lambda_i$ for all $1\leq i\leq m$. For this $X'$, we have $x_N'=x_N$.
\end{Lemma}

\section{Conclusion}
\label{Stream_section_conclusion}

In this paper, we addressed the problem of generating random-bit streams from i.i.d. sources with unknown distributions.
First, we considered the case of biased coins and derived a simple algorithm to generate random-bit streams.
This algorithm achieves the information-theoretic upper bound on efficiency.
We showed that this algorithm can be generalized to generate random-bit streams from an arbitrary $m$-sided die with $m>2$, and its information efficiency is also asymptotically optimal.
Furthermore, we demonstrated that by applying the (generalized) random-stream algorithm, we can
generate random-bit streams from an arbitrary Markov chain very efficiently.

%
%
%
%
%




\end{document}